\documentclass[journal]{IEEEtran}

\usepackage{caption}
\usepackage{titlesec}
\usepackage{etoolbox}
\patchcmd{\thmhead}{(#3)}{#3}{}{}
\usepackage[monochrome]{xcolor}
\usepackage{cite}
\usepackage{interval}
\titleformat{\subsubsection}[runin]{\normalsize\bfseries}{}{}{}
\titlespacing*{\subsubsection}{1pt}{1pt}{3pt}
\ifCLASSINFOpdf
\else
\fi
\usepackage{graphicx}
\usepackage{xcolor}

\usepackage{tabularx,booktabs}
\newcolumntype{C}{>{\centering\arraybackslash}X} 
\usepackage{amsmath}
\usepackage{amssymb}
\usepackage{algorithm}
\usepackage{babel}
\usepackage{algorithmic}
\usepackage{multicol}
\usepackage{subcaption}
\usepackage{chngpage}
\usepackage{array}
\newcolumntype{$}{>{\global\let\currentrowstyle\relax}}
\newcolumntype{^}{>{\currentrowstyle}}
\newcommand{\rowstyle}[1]{\gdef\currentrowstyle{#1}%
  #1\ignorespaces
}

\usepackage{tikz}
\usetikzlibrary{shapes.geometric, arrows}
\tikzstyle{process} = [rectangle, minimum width=3cm, minimum height=1cm, text centered, text width=3cm, draw=black, fill=white!30, scale = 1]
\tikzstyle{arrow} = [thick,->,>=stealth]
\begin{document}
%
\title{Fuzzy Norm-Explicit Product Quantization\\ for Recommender Systems}
%
%
%
\author{~Mohammadreza Jamalifard,~\IEEEmembership{Member,~IEEE,} \and Javier~Andreu-Perez,~\IEEEmembership{Senior Member,~IEEE,} \and Hani~Hagras,~\IEEEmembership{Fellow,~IEEE,} \and Luis~Martínez López,~\IEEEmembership{Senior Member,~IEEE}

\thanks{M. Jamalifard, Hani Hagras \& J. Andreu-Perez are with the Centre for Computational Intelligence at the University of Essex, United Kingdom. J. Andreu-Perez \& L. Martínez López are with group 
$Simbad^{{2}}$ within the Department of Computer Science at the University of Jaén, Spain.}}

\newtheorem{thm}{Theorem}[section]
\newtheorem{lem}[thm]{Lemma}
\newtheorem{proof}{Proof}[section]
\newtheorem{prop}[thm]{Proposition}
\newtheorem{cor}{Corollary}
\newtheorem{defn}{Definition}[section]
\newtheorem{conj}{Conjecture}[section]
\newtheorem{exmp}{Example}[section]
\newtheorem{rem}{Remark}

\maketitle

\begin{abstract}
As the data resources grow, providing recommendations that best meet the demands has become a vital requirement in business and life to overcome the information overload problem. However, building a system suggesting relevant recommendations has always been a point of debate. One of the most cost-efficient techniques in terms of producing relevant recommendations at a low complexity is Product Quantization (PQ). PQ approaches have continued developing in recent years. This system's crucial challenge is improving product quantization performance in terms of recall measures without compromising its complexity. This makes the algorithm suitable for problems that require a greater number of potentially relevant items without disregarding others, at high-speed and low-cost to keep up with traffic. This is the case of online shops where the recommendations for the purpose are important, although customers can be susceptible to scoping other products. A recent approach has been exploiting the notion of norm sub-vectors encoded in product quantizers. This research proposes a fuzzy approach to perform norm-based product quantization. Type-2 Fuzzy sets (T2FSs) define the codebook allowing sub-vectors (T2FSs) to be associated with more than one element of the codebook, and next, its norm calculus is resolved by means of integration. Our method finesses the recall measure up, making the algorithm suitable for problems that require querying at most possible potential relevant items without disregarding others. The proposed approach is tested with three public recommender benchmark datasets and compared against seven PQ approaches for Maximum Inner-Product Search (MIPS). The proposed method outperforms all PQ approaches such as NEQ, PQ, and RQ up to +6\%, +5\%, and +8\% by achieving a recall of 94\%, 69\%, 59\% in Netflix, Audio, Cifar60k datasets, respectively. More and over, computing time and complexity nearly equals the most computationally efficient existing PQ method in the state-of-the-art.\newline
\end{abstract}


%
\IEEEpeerreviewmaketitle

\section{Introduction}
%
%
%
%
\IEEEPARstart{T}{he} nature of recommender systems has a certain characteristic: incremental progress and performance improvement using user experience. Hence, organizations can improve by uninterruptedly learning from their recommenders \cite{lee2020different}.
In practice, it has been reported that about 80\% of what people follow on Netflix is the result of the recommendations, and a combination of penalization and recommendation effect save Netflix more than \$1B annually \cite{ko2022survey}. Similarly, in \cite{deldjoo2022review} is stated that 35\% of amazon purchases come from recommendation lists. The application of recommender systems is not solely limited to commercial purposes, and it has been popular in healthcare decision-making during recent years, among others \cite{tran2021recommender}.\newline\newline 
Model-based approaches have been actively studied for years in the recommender systems field. In this sense, the core of model-based methods is to formulate user-item preference in terms of the inner product of the user and the item's vectors \cite{zhu2018learning}. The most preferred item by a user is predicted to be the one that has the maximum inner product with the user query vector (i.e., MIPS). However, an exhaustive search through the inner product space of billion vectors in the largest dataset due to its high complexity of both space and time and thereby costs (e.g., increasing processing units, cloud computing, semiconductors) brings up the necessity of a technique to increase the computational efficiency \cite{le2021efficient}.
By constructing a linear transformation, MIPS is redefined to the Nearest Neighbor Search problem, and this is where product quantization is proved to come through the task of cutting down largely the computational cost \cite{gholami2021survey}. To reduce the norm error in Product Quantization, the Norm-explicit Quantization technique has been proposed, which has proven to be successful in providing a better estimation of MIPS than other vector quantization methods such as PQ, Optimized PQ (OPQ), RQ, and AQ in terms of recall as the measure of performance. Besides, this method has proven to be efficient with respect to recall even with large quantization errors and/or less number of codebooks in comparison to its counterparts because quantizing the norm of items reduces the norm error, which leads to better performance.

\subsubsection*{Why MIPS-based recommender systems?}\label{subsec:1.1}
 There are varied ways to build a recommender system, either as a collaborative filter or content-based, modeling via matrix factorization to neural networks;\cite{bobadilla2013recommender}\cite{yera2017fuzzy}\cite{lu2015recommender}\cite{perez2021content} each approach has its cons and pros. Choosing a method, in this case, could be highly dependent on the nature of the problem. MIPS is a well-established method for recommender systems that are computationally efficient and scalable to large multidimensional datasets \cite{al2021survey}. It is also proven efficient in many recommendation tasks, yielding fast search time. It can be combined with other algorithms in frameworks to yield higher or improved performance. In business systems, it is frequent that performance requirement standards such as time, space complexity, and amount of relevant items successfully retrieved are factored in. Deep neural networks mark the ceiling of precise recommendations in all benchmarks. Still, their computational complexity and minimum hardware requirements could spiral costs if a business aims to scale for the highest customer traffic possible to ensure more sales and revenues. 
\subsubsection*{What is norm-explicit product quantization (NEQ)?}\label{subsec:1.2} Product Quantization is an efficient baseline method in terms of time and space complexity to address the nearest search problem. NEQ is its upgraded version \cite{dai2020norm}. This approach exploits that the norm is essential for MIPS because it is used in calculating the inner product, which is the basis of the similarity of the vectors. Further, the vector set is quantized into an approximated sub-vector forming a codeword of a codebook, similar to a divide and conquer or data compression strategy, which permits encoding of the norm separately and reduces the error. However, the association of a vector with a codeword is a hard association, which can impact the estimation of query vectors within the codebook's boundaries. Therefore, the objective is to integrate fuzziness in NEQ by making the basis of computation in the codebook a collection of T2FSs whose boundaries are defined via soft clustering. The codebooks, as well as being computationally cost-effective in the stage of codebook construction\cite{wu2019vector}, are formed of T2FSs (codewords) to implement quantization, and vectors have more than one soft association with several codewords. 
\subsubsection*{The highlights and findings of this work are:}\label{subsec:1.3} 
\begin{itemize}

\item Proposing a novel method based on a fuzzy type-2 clustering algorithm to construct the fuzzy codebooks and then fuse the results of the fuzzy codebooks using Sugeno Integral to implement norm-explicit quantization.
\item This method can be run on a large set of data with a surprisingly low running time and high performance measured by the recall to indicate the relevance of the output.
\item The proposed approach improves the recall per item rate in comparison to Norm-Explicit Product Quantization and other similar approaches such as PQ or OPQ on Netflix, Audio and Cifar60k datasets.
\end{itemize}

Next, section \ref{sec:related work} provides a related comparative work to the proposed method. Section \ref{sec:background} provides an overview and preliminaries about the recommender system, Nearest Neighbour Search (NNS), and MIPS, its different types, and recent relevant research on the subject. The following section \ref{sec:proposed} describes Fuzzy-2 NEQ in every detail. Subsequently, the results and comparison of our method with baseline methods are presented in section \ref{sec:results}.  Finally, section \ref{sec:conclusion} presents a conclusion of the research and questions for further studies.

\section{Related Work}\label{sec:related work}
In this section, a number of related research works \textcolor{blue}{are} reviewed to give a better picture of what has already been done. Jegou et al. introduced PQ, which defines $M$ sub-dataset out of the original dataset with a size of $d/M$ features. In this method, K-means is used to train the codebooks of each dataset independently. A codeword in this approach would be a $d/M$-dimensional vector. This way, the approximated value of $x$ is a fusion of its corresponding codewords \cite{zhong2022evaluating}.
OPQ uses an orthonormal matrix $R$ to rotate the items by $Rx$ before using PQ. OPQ leads to lower quantization error when either the
features are correlated, or some features show a larger value of variance
in comparison to others. However, the codebook training phase in OPQ is slightly more complex as it contains multiple rounds of alternating optimization of the codebooks and the rotation matrix $R$ \cite{guo2022semantic}. In 2010 Chen et al. suggested the RQ approach in which each codebook covers whole features, and each codeword is a $d$-dimensional vector. The original data are used to train the first codebook with K-means, and the concept of residues is used to train the second codebook. The training phase is carried out recursively, which means $m$-th codebook is trained based on the residues from ($m-1$) previous codebooks \cite{fan2022comprehensive}. \textcolor{blue}{Babenko et al.} introduced AQ, which collectively improved RQ by optimizing all $M$ codebooks. In this approach, they use Beam search \cite{meister2020if} to do encoding of the codebook (e.g., finding the optimal codeword indexes of an item within the codebooks) and a least-square relation to optimize the codebooks subject to the proposed encoding \cite{fan2022comprehensive}.
In 2020, Dai and his colleagues tried using signal compression like other vector quantization-based methods \cite{boaventura2020shape}. Applying this idea which separately quantizes the magnitude and direction of a signal to have a fair level of efficiency, leads to a decrease in performance. However, the Norm-Explicit approach quantizes direction and norm separately, resulting in a more efficient performance for the problem of maximum inner product search (MIPS) \cite{dai2020norm}. Quantizing the norm has also been claimed to reduce the quantization error while the dynamic range of data is rather large\cite{kairouz2021advances}. Norm-Explicit Product Quantization has signified advancements in terms of performance in comparison to the methods discussed above. Nevertheless, it relies on discrete quantization. If we remove this constraint, we can have a method that can operate with a less restricting and smoother definition of the codebook, augmenting its quality.
\newline
\section{Background}\label{sec:background} 
To start with this section, a more or less formal definition of the recommender system is worth revising. Briefly speaking, a recommender system, or a recommendation system (the word 'system' may be interchangeable with similar words like 'engine' or even 'platform'), is a subclass of information filtering system that delivers suggestions for items that are most to the point to a specific use. Elaine Rich introduced the idea of a recommender system in 1979 \cite{jannach2021survey}. She looked for a way to recommend a user a book they might be interested in. Her idea was to make a system that asks a particular user question and assigns them stereotypes subject to their responses. Depending on a user's stereotype, they would then get a recommendation for a book they might like \cite{ricci2022recommender}. In recommender systems, two problems convertible to one another need to be discussed. The first is NNS, and the other is Maximum Inner Product Search (MIPS).
\subsection{Recommender Systems, NNS and MIPS}\label{subsec:2.1}
Recommender systems are used in a wide range of applications, ranging from playlist generators for video and music services, product recommenders for online shops, content recommenders for social media platforms, and open web content recommenders \cite{sanz2019information}.
Recommender systems have also been developed to explore research articles \cite{nikzad2019state} collaborators \cite{gonccalves2019automated}, financial services, \cite{iovine2023virtual} and predicting cancer drug response \cite{suphavilai2018predicting}. These systems usually employ collaborative filtering and/or content-based filtering method separately or together \cite{haubner2020applying}. Besides, they might use similar systems like knowledge-based systems. Collaborative filtering approaches build a model based on a user's previous behaviours (items previously bought or chosen and/or ratings given to those items), and analogous decisions made by other users could be a source of information for this approach \cite{yera2017fuzzy}. This model is then utilized to predict items or ratings of the item that the user may be inclined to \cite{singh2020scalability}. Content-based filtering approaches use a set of discrete, pre-tagged attributes of an item to recommend additional items with similar properties \cite{zhang2020survey}.
Each type of system has its advantages and disadvantages. Collaborative filtering may need a great deal of information to start with, which is an example of the cold start problem, and is a frequent topic in collaborative filtering research\cite{elahi2016survey}\cite{mosqueira2022human}. In contrast, content-based filtering can be done by having a limited scope of information and without information about user behaviours, which users may not be willing to share. Fast recommendations can enhance user experience when user behavior data is not available. Enabling rapid prototyping, users could quickly and actively find what they are looking through engaging with an interactive website or application \cite{subramonyam2022UX}.\newline\newline
 A fast way for finding the preferred items for a specific user is by searching through the inner products of items by user query vectors and finding vectors with the maximum value of the inner product. Maximum Inner Product Search (MIPS) using a simple manipulation could be viewed equivalently to Nearest Neighbor Search (NNS) in which maximization of inner product in MIPS is considered as the same as distance minimization in NSS \cite{gholami2021survey}. 
The fundamental relations of MIPS and NNS are as follows:
\begin{equation}
   x = argmin_{\boldsymbol{x}_i \in \boldsymbol{D}} \lvert |\boldsymbol{x}_i - \boldsymbol{q}| \rvert_2
\end{equation}

\begin{equation}
    x = argmax_{\boldsymbol{x}_i \in \boldsymbol{D}} \boldsymbol{x}_{i}^T \boldsymbol{q}
\end{equation}
In which $\boldsymbol{D}$ is frequently a massive dataset queried from, $x_{i}$ is a database vector, $\boldsymbol{q}$ is the given query, $\lvert|.|\rvert_2$ is L2-norm, and $x$ would be the nearest neighbor for the query $\boldsymbol{q}$ or maximum inner product value identically using a simple trick or more specifically a linear transformation.
Utilizing the following relations facilitates the construction of the transformation:
\begin{equation}
    \phi = max_{\boldsymbol{x}_i \in \boldsymbol{D}} \lvert |\boldsymbol{x}_i| \rvert_2
    \label{equ:3}
\end{equation}

\begin{equation}
    \boldsymbol{z}_i = (\sqrt{\phi^2 - \lvert |\boldsymbol{x}_i| \rvert_2^2}, \boldsymbol{x}_{i}^T)^T
    \label{equ:4}
\end{equation}

\begin{equation}
    \boldsymbol{q}_z = (0, \boldsymbol{q}^T)^T
    \label{equ:5}
\end{equation}
as it is observed, $\boldsymbol{z}_i$ and $\boldsymbol{q}_z$ are modified item and query vectors respectively, and $\phi$ represents the maximum norm of items in the dataset \cite{le2021efficient}. Using the relations (\ref{equ:3})-(\ref{equ:5}), we have the following theorems with proofs \cite{bachrach2014speeding}:
\begin{thm}
For every $\boldsymbol{z}_i$, $\lvert |\boldsymbol{z}_i|\rvert^2_2 = \phi$
\end{thm}
\begin{proof}
Using the relation (4), we have $\lvert|\boldsymbol{z}_i|\rvert^2_2 = \lvert |\boldsymbol{x}_i| \rvert^2_2 - \lvert|\boldsymbol{x}_i| \rvert^2_2 + \phi$ $\square$
\end{proof}

\begin{thm}
For every $\boldsymbol{x}_i$, $\boldsymbol{z}_i$ we have $\boldsymbol{z}_{i}^T\boldsymbol{q}_{z} = \boldsymbol{x}_{i}^T\boldsymbol{q}$
\end{thm}

\begin{proof}
Since the first component of $\boldsymbol{q}_z$ equals zero, we know that $(w^T)^T = w$ for an arbitrary vector $\boldsymbol{w}$. Hence, we have $(\sqrt{\phi^2 - \lvert |\boldsymbol{x}_i| \rvert_2^2}, \boldsymbol{x}_{i}^T).(\boldsymbol{0}, \boldsymbol{q}^T)^T = \boldsymbol{0} + \boldsymbol{x}_i^T(\boldsymbol{q}^T)^T = \boldsymbol{x}_i^T\boldsymbol{q}$.\\
Consequently,
$\boldsymbol{z}_i^T\boldsymbol{q}_z = \boldsymbol{x}_i^T\boldsymbol{q}$ $\square$

\end{proof}

Having the mentioned relations, the linear transformation based on the inner product term is simply as follows:
\begin{equation}
    \lvert |\boldsymbol{q}_z - \boldsymbol{z}_i| \rvert^2_2 = \lvert|{\boldsymbol{q}_z}|\rvert^2_2 - 2\boldsymbol{z}_i\boldsymbol{q}_i^T + \lvert |\boldsymbol{z}_i|\rvert^2_2 = \lvert|\boldsymbol{q}|\rvert^2_2 + \phi^2 - 2\boldsymbol{x}_i^T\boldsymbol{q} 
\end{equation}
relation (6) shows the distance relation in NNS can be written in terms of the inner product of item and query vectors which is the transformation of interest to make the nearest neighbor search problem identical to MIPS \cite{bachrach2014speeding}.
\subsection{Product Quantization}
As a general concept, Product Quantization (PQ) is a vector quantization approach representing high dimensional vectors as a Cartesian product of sub-spaces for quantizing them separately. Before explaining the basic idea of product quantization, we provide some insight into the structures of this method. From a mathematical point of view, the definition of a quantizer \cite{wu2019vector} is well-worth discussing.
\begin{defn}[Quantizer]
a quantizer is a function $q$ mapping a $D$-dimensional vector $\boldsymbol{x} \in \mathbb{R}^D$ to a vector $\boldsymbol{q(x)} \in \mathcal{C}=\{\boldsymbol{c}_i ; i \in \mathcal{I}\}$, where the index set defines as $\mathcal{I}$ for simplicity considered to be finite: $\mathcal{I} = 0,.., k-1$. 
\end{defn}

We have used a fuzzy clustering technique called Interval Type-2 Fuzzy Possibilistic C-means, which has been introduced in \cite{rubio2016interval} to add fuzzy quality to the process of the algorithm.

Using weights as exponents, this algorithm uses $\xi$ and $\eta$ to describe the fuzziness and possibility, representing an interval or simply a range rather than an exact value \cite{mendel2017interval}.
The choice of the best parameters for the fuzzy interval is carried out by using Genetic Algorithm\cite{alibrahim2021hyperparameter} (a color-map of grid values is added as an appendix). The best set of values for $\xi_{1}, \xi_{2}$ are 8.5, 9.1 respectively.
\section{Fuzzy-2 NEQ}\label{sec:proposed}
Vector quantization, as one of the data compression methods, could largely reduce the cost of MIPS by encoding the input into a lower dimensional subspace using codebooks. 
An ordinary vector quantization may have issues such as the high-running time by growth in the number of inputs or space complexity challenges to store many records for a $D-$dimensional vector.
\subsection{Codebook Definition}
The reproduction values $\boldsymbol{c}_i$ mentioned in the background section are normally called centroids, and it would be the initial notion to define codebooks effortlessly as below:
\begin{defn}[Codebook]
The set of reproduction values $\mathcal{C}$ is, in fact, the codebook of size $k$.
\end{defn}
But how our model as fuzzy-2 NEQ would be convergent as a quantization method needs to be satisfied Lloyd's properties \cite{levrard2018quantization}.
In quantization, the input space is partitioned into a set of convex regions called Voronoi cells.
\begin{defn}[Voronoi Cell]
The set $\mathcal{V}_i$ of vectors mapped to a given index $i$ called as a Voronoi Cell defines as following:
\begin{equation}
\mathcal{{V}}_{i} \triangleq\left\{\boldsymbol{x} \in \mathbb{R}^{D}: q(\boldsymbol{x})=\boldsymbol{c}_{i}\right\} .
\end{equation}
\end{defn}
the Lloyd's optimality properties are as follows:

\begin{equation}q(\boldsymbol{x})=\arg \min _{\boldsymbol{c}_{i} \in \mathcal{C}} d\left(\boldsymbol{x}, \boldsymbol{c}_{i}\right) \end{equation}
\begin{equation}\boldsymbol{c}_{i}=\mathbb{E}_{X}[\boldsymbol{x} \mid i]=\int_{{V}_{i}} p(\boldsymbol{x}) \boldsymbol{x} d \boldsymbol{x} \end{equation}
the first condition simply states that each reproduction value $\boldsymbol{c}_i$ is assigned to its nearest data point. the second condition points out that the value of $\boldsymbol{c}_i$ must be the expected value of the vectors in Voronoi Cell of index $i$ (e.g., $\mathcal{V}_i$).
Interval Type-2 Fuzzy Possibilistic C-means comes up with a near-optimal codebook that is less susceptible to noise than other variations of fuzzy c-means\cite{rubio2016interval}. It works by assigning vectors to centroids to meet the optimality criteria. Then, we fuse the result of the fuzzy clustering algorithm, a set of nearly-optimal codebooks using the Sugeno integral since Sugeno integral is an ordinal aggregation method which can grade similarity and in the case of our problem, it is used for fusing the set of codebooks as the output of the fuzzy clustering method to a single crisp codebook. Besides, adding a stop condition of improvement at the 7th step of the algorithm helps us to have at least a local optimum result.
In the product quantization case, the proposed issues are settled by selecting the number of components that should be quantized separately. To clarify by an example, an input vector denoted by $\boldsymbol{x}$ is split into $m$ distinct sub-vectors $\boldsymbol{o}_j$ of dimension $D^* = D/m$ where $m$ is a divisor of $D$ and $1\leq j \leq m$. The sub-vectors are quantized separately using $m$ distinct quantizers. Dividing the primary vector into sub-vectors provides two significant advantages. Firstly, it aids in preserving the local structure of the data. Secondly, it facilitates dimensionality reduction. By segmenting the vector into smaller elements, we can better comprehend the inherent features and correlations within the data. This offers a computational advantage by reducing the complexity of the problem. Since each sub-vector is separately quantized using its own quantizer, it enables customized quantization for each sub-vector. Therefore, the quantization process is tailored to the distinct features and distribution of every sub-vector. The selection of the number of sub-vectors ($m$) in product quantization impacts the balance between accuracy and complexity. A large $m$ helps in seizing more detailed information but simultaneously boosts computational complexity. Conversely, utilizing smaller sub-vectors more efficiently captures local information. However, overly small sub-vectors might lead to a loss of global information. Subsequently, the vector $\boldsymbol{x}$ which has been chopped into $m$ number of $D^*-$dimensional vectors would be mapped into a vector $\boldsymbol{q(x)}$ using the quantizer function as below:
\begin{equation}
    \underbrace{\boldsymbol{x}_1,...,\boldsymbol{x}_{D^*}}_\text{$\boldsymbol{o}_1$}, ...\,, \underbrace{\boldsymbol{x}_{D-D^* + 1}, ..., \boldsymbol{x}_D}_\text{$\boldsymbol{o}_m$} \mapsto{q_1(\boldsymbol{o}_1(\boldsymbol{x})), ...\,, q_m(\boldsymbol{o}_m(\boldsymbol{x}))}
\end{equation}
where $q_j$ is a low-complexity quantizer associated with
the $j$-th sub-vector. With the sub-quantizer $q_j$,
the index set $\mathcal{I}_j$, the codebook $\mathcal{C}_j$ and the corresponding reproduction values $c_{j,i}$ are being associated.
A reproduction value of the product quantizer is identified by an element of the product index set $\mathcal{I}=\mathcal{I}_1\times...\times\mathcal{I}_m$ The codebook is therefore defined as the Cartesian product of $\mathcal{C} = \mathcal{C}_1 \times, ..., \times \mathcal{C}_m$ and a centroid of this set is the concatenation of centroids of the $m$ sub-quantizers. Based on the assumption that all sub-quantizers share the same finite number $k^*$ of reproduction values, the total number of centroids yielded by $k = (k^*)^m$. Then the number of codebooks is $m$ containing $k^*$ codewords of length $D^*$. In order to learn the codebooks, a clustering technique is employed on each of them separately, and then as has been mentioned earlier, estimated item $\boldsymbol{x}$ denoted by $\boldsymbol{\dot{x}}$ worked out by concatenation of its corresponding $D^*$ dimensional codewords \cite{dai2020norm}. 
\subsection{Fuzzy Norm-Explicit Quantization}
\begin{figure}[!ht]
\centering
\begin{tikzpicture}[scale=0.77, transform shape, node distance=1.37cm]
\node (p1) [process, yshift = -0.5cm]{Loading The Processed Data};
\node (p2) [process, right of = p1, xshift =3.5cm]{Applying PQ algorithm by using the Fuzzy Type-2 Clustering to Construct Vector Codebooks};
\node (p3) [process, below of = p2, yshift = -2cm]{Fusing The Codebooks By Taking Sugeno Integral From The Result};
\node (p4) [process, below of = p1, yshift = -2cm]{Computing An Estimation Of The Direction Vector Based On The Constructed Codebooks};
\node (p6) [process, below of = p3, yshift = -2cm]{Constructing The Norm Codebooks To Quantize The Relative Norm};
\node (p5) [process, below of = p4, yshift = -2cm]{Getting The Relative Norm Of The Item "$\boldsymbol{x}$"};
\node (p7) [process, below of = p6, xshift = -2.5cm, yshift = -0.5cm]{Returning Both The Norm and Vector Codebooks};
\draw [arrow] (p1) -- (p2);
\draw [arrow] (p2) -- (p3);
\draw [arrow] (p3) -- (p4);
\draw [arrow] (p4) -- (p5);
\draw [arrow] (p5) -- (p6);
\draw [arrow] (p6) -- (p7);
\end{tikzpicture}
\caption{The figure above represents the stages of the algorithm}\label{fig:tikz}
\end{figure}
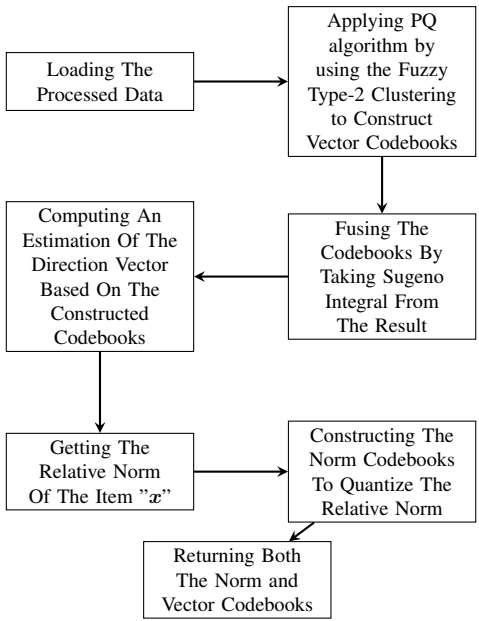

To improve the performance of product quantization, the point of using norm values has been introduced as a complementary idea conjunctive to the notion of PQ to reduce the error of quantization and, consequently, improve the performance of product quantization \cite{dai2020norm}. MIPS would take advantage of methods that explicitly cut down the error rate in the norm since an accurate norm is significant to MIPS performance. Hence, the core of Norm-Explicit quantization is to quantize the norm $\lvert |\boldsymbol{x}| \rvert_2$ and the direction vector, which is unit vector $\boldsymbol{x}$ of the items distinctively. The norm is encoded explicitly using separate codebooks to achieve a small error, while the direction
vector can be quantized using an existing vector quantization approach without making notable changes. \newline\newline In detail, the $m$ codebooks in NEQ are split into two parts. \textcolor{blue}{The first part, referred to as $m'$, consists of codebooks that are considered as norm codebooks}, denoted by $\mathcal{L}_p$ for $p \in \{1,...,m'\}$ in which each codeword $\boldsymbol{{l}}_p[n] \in \mathbb{R}$ for $1 \leq n \leq k^*$.
The other $m-m'$ codebooks $\boldsymbol{c}_{m'+1}, \boldsymbol{c}_{m'+2} ,..., \boldsymbol{c}_m$ are vector codebooks for the direction vector. To calculate $\dot{x}$ in this method, the following formula is applicable (further detail be consulted in \cite{dai2020norm}):
\begin{equation}
    \boldsymbol{\dot{x}} = \left(\sum_{m=1}^{m'} {\boldsymbol{l}}_m[i_x^m]\right)\cdot\left(\sum_{m=m'+1}^{m} \boldsymbol{c}_m[i_x^m]\right)
\end{equation}
in which $i_x^m$ represents the $m$-th codeword index of the item $x$. the inner product introduced in relation (16) can be worked out by Algorithm 1 using $\boldsymbol{\dot{x}}$, which is the approximation of the item $\boldsymbol{x}$ indeed.
\begin{algorithm}[h]
\caption{Estimation of Inner Product Algorithm}\label{alg:cap}
\begin{algorithmic}[1]
\REQUIRE Query $\boldsymbol{q}$, $m$ Codewords of the item $x$ e.g $i_x^1,...,i_x^m$
\ENSURE An estimation of $\boldsymbol{q}^T\boldsymbol{x}$
\STATE $\boldsymbol{l}=\boldsymbol{0}, \boldsymbol{r}=\boldsymbol{0}$
\WHILE{$s=1$ to $m'$}
\STATE $\boldsymbol{l} = \boldsymbol{l} + \boldsymbol{l}_s[i_x^s]$
\STATE $s=s+1$
\ENDWHILE
\WHILE{$s=m'+1$ to $s=m$}
\STATE $\boldsymbol{r} = \boldsymbol{r} + \boldsymbol{q}^T\boldsymbol{c}_s[i_x^s]$
\STATE $s = s + 1$
\ENDWHILE
\STATE return $\boldsymbol{l} \cdot \boldsymbol{r}$
\end{algorithmic}
\end{algorithm}
\begin{algorithm}[t]
\caption{Codebook Training}\label{alg:cap2}
\begin{algorithmic}[1]
\REQUIRE dataset $\boldsymbol{D}$, number of codebooks $m$, number of norm codebooks $m'$ 
\ENSURE $m'$ trained codebooks of the norms, $m-m'$ trained vector codebooks
\STATE Compute the direction vector $\boldsymbol{x'} = \frac{\boldsymbol{x}}{\lvert |\boldsymbol{x}|\rvert_2}$
\STATE Train $m-m'$ vectors using the type-2 fuzzy clustering approach to build set of fuzzy codebook ${\mathcal{\tilde {C}}}$
\STATE Aggregate the fuzzy codebooks generated in the previous step using Sugeno integral
\STATE Encode $\boldsymbol{x'}$ by the vector codebooks, obtain an estimation of $\boldsymbol{\bar{x}}$ of $\boldsymbol{x'}$ based on the codebooks
\STATE Compute relative norm by $\frac{\lvert |\boldsymbol{x}|\rvert}{\lvert |\bar{\boldsymbol{x}}| \rvert}$
\STATE Train $m'$ norm codebooks to quantize relative norms
\STATE Return $m$ codebooks
\end{algorithmic}
\end{algorithm}
The next step to solving the problem in the big picture is to train vector codebooks and norm ones. The second and third steps of algorithm 2 generate fuzzy codebooks and then fuse the results using the Sugeno integral to obtain a crisp codebook so it becomes consistent with the next stage of the algorithm. After discussing vector quantization and, more specifically, NEQ, we will get to the stage of employing a fuzzy approach to carry out the vector quantization task specified at the second step of Algorithm 2. \newline\newline Although NEQ has been shown to improve existing vector quantization method in terms of performance, having a fuzzy view in this problem is worth considering due to nature of large and sparse datasets which increases the likelihood of datum in overlapping clusters. Instead of using hard clustering to train the vectors (e.g., K-means), we employed a soft Interval type-2 Fuzzy Possibilistic C-means clustering to train the codebooks. As a short explanation of this fuzzy clustering algorithm, it consists of an expansion of Fuzzy Possibilistic C-Means employing a Fuzzy Type-2 approach \cite{rubio2016interval}. The next step would be training the codebooks using the proposed fuzzy clustering algorithm to ensure we used the property of fuzziness, which considers soft margins to the clusters and overlapping clusters. However, there is a challenge when a fuzzy algorithm constructs codebooks, and the challenge would be the structure of the fuzzy output, which is a tensor for each cluster. Having a fuzzy codebook vector as an output on hand compels us to think of an idea to transform the tensor made up of fuzzy codebook vectors into a consistent input for the next stage of the algorithm. Therefore, to use Algorithm 2, having a crisp result is necessary for setting the next steps that can be done through aggregation. Hence, the codebooks would train using a type-2 fuzzy approach resulting in a fuzzy output. To achieve a result which is crisp, aggregation would be significant. In this case, to fuse the resulted data, we decided to use Sugeno Integral \textcolor{blue}{because of its non-linear integration capability\cite{klement1983nonlinearity}, and showing better performance than Choquet integral as a similar aggregation function in our case as well as its nature of weighted aggregation\cite{rico2019qualitative}, which can be crucial for handling the complex and non-linear relationships typical in high-dimensional recommender system data. Sugeno Integral} mathematically defines over the fuzzy set of the codebooks ${\displaystyle {\mathcal{\tilde {C}}}}$ of the function ${\displaystyle h}$ h refers to cluster centers for the fuzzy measure ${\displaystyle g}$ as:\newline
\begin{equation}
\int_\mathcal{\tilde{C}} h(x) \circ g = \int_\mathcal{X} \left[h_\mathcal{\tilde{C}}(x) \wedge h(x)\right] \circ g 
\end{equation}
\newline
where ${\displaystyle h_{\mathcal{\tilde{C}}}(x)}$ is the membership function of the fuzzy set of codebooks ${\mathcal{\tilde {C}}}$, and $\mathcal{X}$ is the domain of the function $h$ \cite{bardozzo2021sugeno}. In addition, the product has been used as a t-norm of Sugeno integral. Product quantization involves dividing a high-dimensional space into smaller subspaces and quantizing each of these subspaces separately.  We propose that each subspace has its own codebook, which is utilized to encode the data. Our approach employs the Sugeno Integral during the fuzzy codebook aggregation phase to merge the fuzzy memberships of the codewords from different fuzzy codebooks. This is accomplished by computing the weighted average of inputs based on their membership values and predetermined weights. The aim is to combine the fuzzy memberships of the codewords, taking into account their respective weights, in order to produce a final output that is a composite of all the individual inputs.\newline\newline The aggregation approach in our model serves the aggregation inference phase and yields a crisp and compatible result for the algorithm's next step, which is encoding the vector codebooks\cite{gagolewski2022data}. Finally, The resulting codebooks that are produced consist of items that have been fragmented and compressed and subsequently arranged in order to identify the top 20 preferred items which are used for the purpose of recommendation. \textcolor{blue}{As the type-2 fuzzy set is capable of dealing with higher level of uncertainty in comparison to the type-1 fuzzy set, it could excel in handling the inherent uncertainty and fuzziness in user preferences, providing a more detailed and accurate representation of user behavior which this can lead to improved personalization in recommendations. To exemplify, we can consider a scenario where a user's interests are not sharply defined but rather fuzzy. For example, if a user shows a preference for both action and drama movies, but with varying degrees of interest, their vector can be associated with codewords representing both categories, but with different membership degrees. This fuzzy association allows for a more elaborate and accurate representation of user preferences and item features. In such cases, a rigid association between item/user vectors and codewords may not capture the details of user preferences or item features effectively.} As the performance measures, recall and running time are proposed in the results section.
As materials, we have used Netflix, Cifar60k, and Audio datasets to benchmark our method. The method's parameter setting of the method in addition to statistical relations of the metrics will be provided in the next section. Besides, we compare the fuzzified Norm-Explicit Product Quantization and its non-fuzzy version to see how much adding a fuzzy approach to the construction of vector codebooks would impact the performance of Norm-Explicit PQ. 
Figure \ref{fig:tikz} describes the general stages of the method in a step-wise fashion.\newline\newline
\\
To generate recommendations using the recommender system, we start by selecting the top-$k$ recommendations—taking, for example, $k=20$ in our case. We utilize encoded representations created by the codebooks for this purpose. The next step involves computing the similarity measure (e.g., Euclidean distance) between the user's encoded representation and the encoded representations of all items. Based on this measure, we rank the items and select the top-20 items as recommendations for the user.

\section{Results}\label{sec:results}
In this section, we try to provide the result of the method discussed in the previous chapter based on metrics such as recall and overall time. Custom Python code was elaborated for implementing methods and the Fancy-Aggregations Python library was utilized to compute the Sugeno integral. \cite{fumanal2021motor}. We employed three well-known datasets, namely Netflix, Audio, and Cifar60k. A summary of their statistics has been provided in Table 1. The Netflix dataset contains user ratings for items, and the researchers applied an ALS-based matrix factorization approach to obtain embeddings for both the items and users. The item embeddings were used as dataset items, while the user embeddings were used as queries. The Cifar60k dataset comprises 6000 32x32 colored images that belong to 10 different categories. The Audio dataset contains recorded voices of men and women in various qualities that range from 512 to 1024 kbps. Lastly, we will also discuss scenarios where our method can be suitable.

\subsection{The Metrics of Performance}\label{sec:4.1}
Running time and recall metrics in the case of our problem can be treated in the context of information retrieval and a type of time complexity in computer systems, in which recall is the number of correct results divided by the number of results that should have been returned \cite{borgman2021effective}.  Recall would be estimated as follows:
\begin{equation}\label{rec}
    Recall = \frac{|Relevant\,Documents \cap Retrieved\,Documents|}{|Relevant\, Documents|}
\end{equation}
and as another metric of performance for the algorithm, precision would be considered which is defined as below:
\begin{equation}\label{prec}
    Precision = \frac{|Relevant\,Documents \cap Retrieved Documents\,|}{|All \, Relevant\, Documents|}
\end{equation}
by having precision and recall, F1-score can be easily computed by the following relation \cite{ricci2021challenges}:
\begin{equation}\label{f1sc}
    F1-score = \frac{2\times Precision \times Recall}{(Precision + Recall)}
\end{equation}
To calculate the running time, we used the time difference between the prior and post-running times of the main function of the code containing algorithms of study. The hardware specification of the system running the algorithm has been added as the footnote.\footnote{ Intel Core i7-8750H 2,2GHz
RAM: 8GB DDR4-SDRAM}. In addition, a short explanation of the asymptotic analysis and comparison of vector quantization methods with other methods is delivered in the following subsections. The relation can be written as:
\begin{equation}
    Running Time = |Start\,Time- Termination\,Time|
\end{equation}
Cifar60k, Audio, and Netflix datasets have been used as the experiment material to measure the method's performance by the recommended metrics above \cite{chatzimparmpas2021empirical}. An overview of these datasets has been provided in a table further in this section.
\subsection {Fuzzy-2 NEQ vs. Baseline Models}
To have a more analytical evaluation of our model, comparing it to similar models is informative. A more detailed discussion of the some baseline methods, including PQ, OPQ, and RQ, has been delivered in related work as a part of the literature review. However, a short description each baseline methods is proposed further in this section. In detail, each model was tested by keeping the number of codebooks as low as possible(e.g., 8) and several clusters 16, 64 to see how efficiently productive these models could be in comparison to Fuzzy-2 NEQ based on their recall/item values.
\begin{figure}
    \centering
    \includegraphics[scale=0.061]{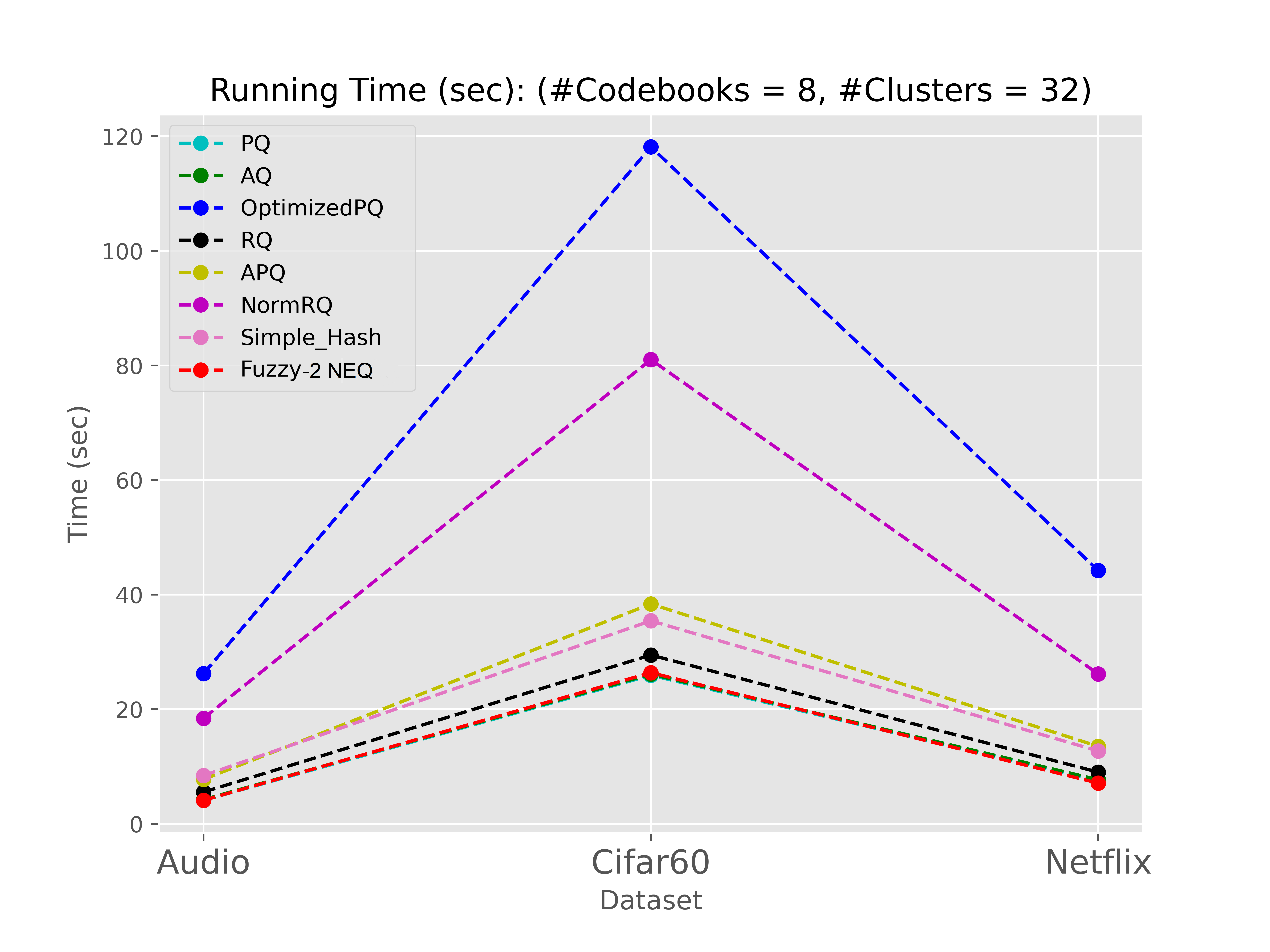}
    \caption{Time comparison of proposed the fuzzy method vs Other Baseline methods}
    \label{fig:timep}
\end{figure}
\begin{figure*}[!htb]
\minipage{0.33\textwidth}
  \includegraphics[width=\linewidth]{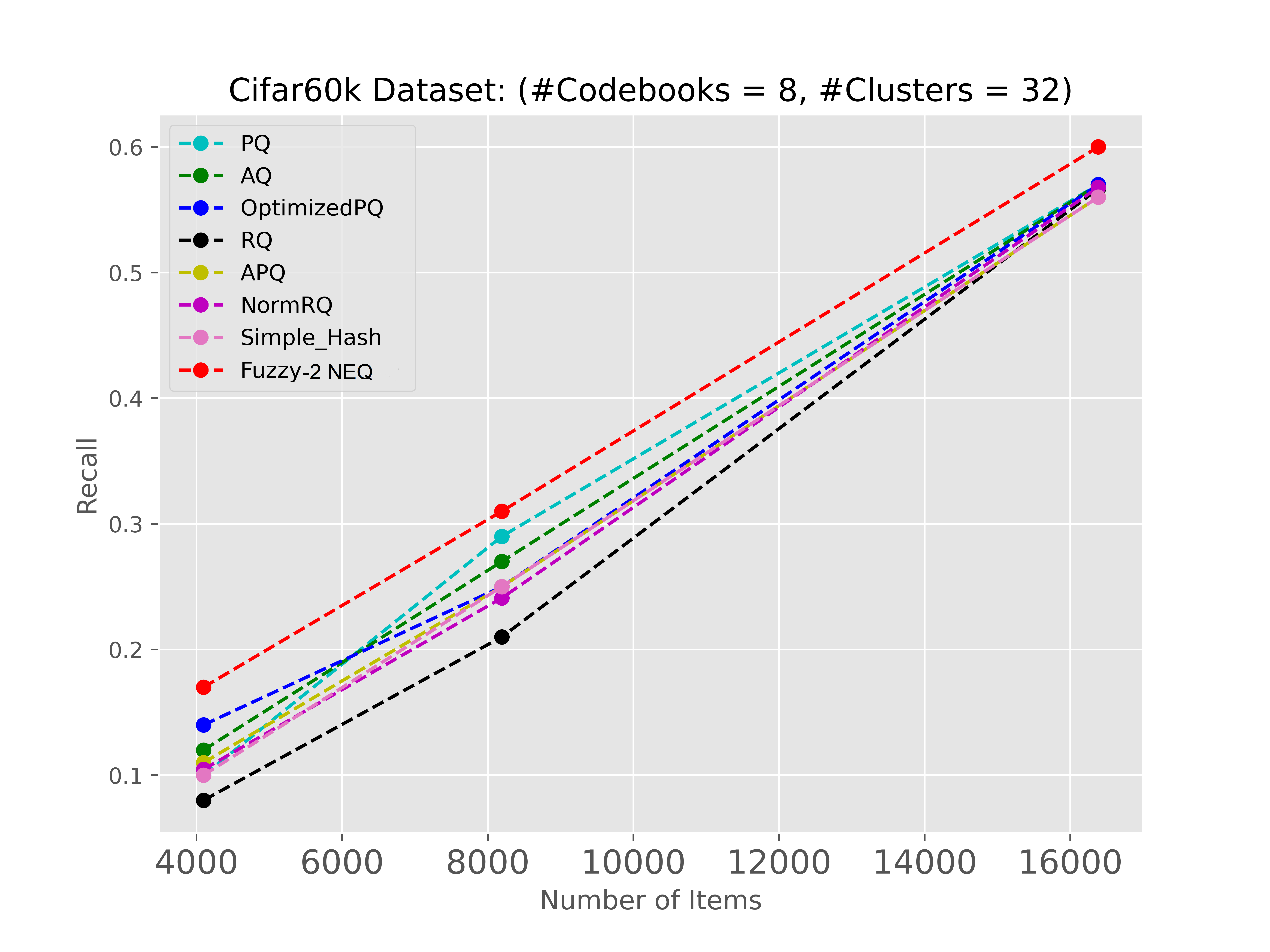}
  \caption{Cifar60k dataset comparison plot}\label{fig:awesome_image1}
\endminipage\hfill
\minipage{0.33\textwidth}
  \includegraphics[width=\linewidth]{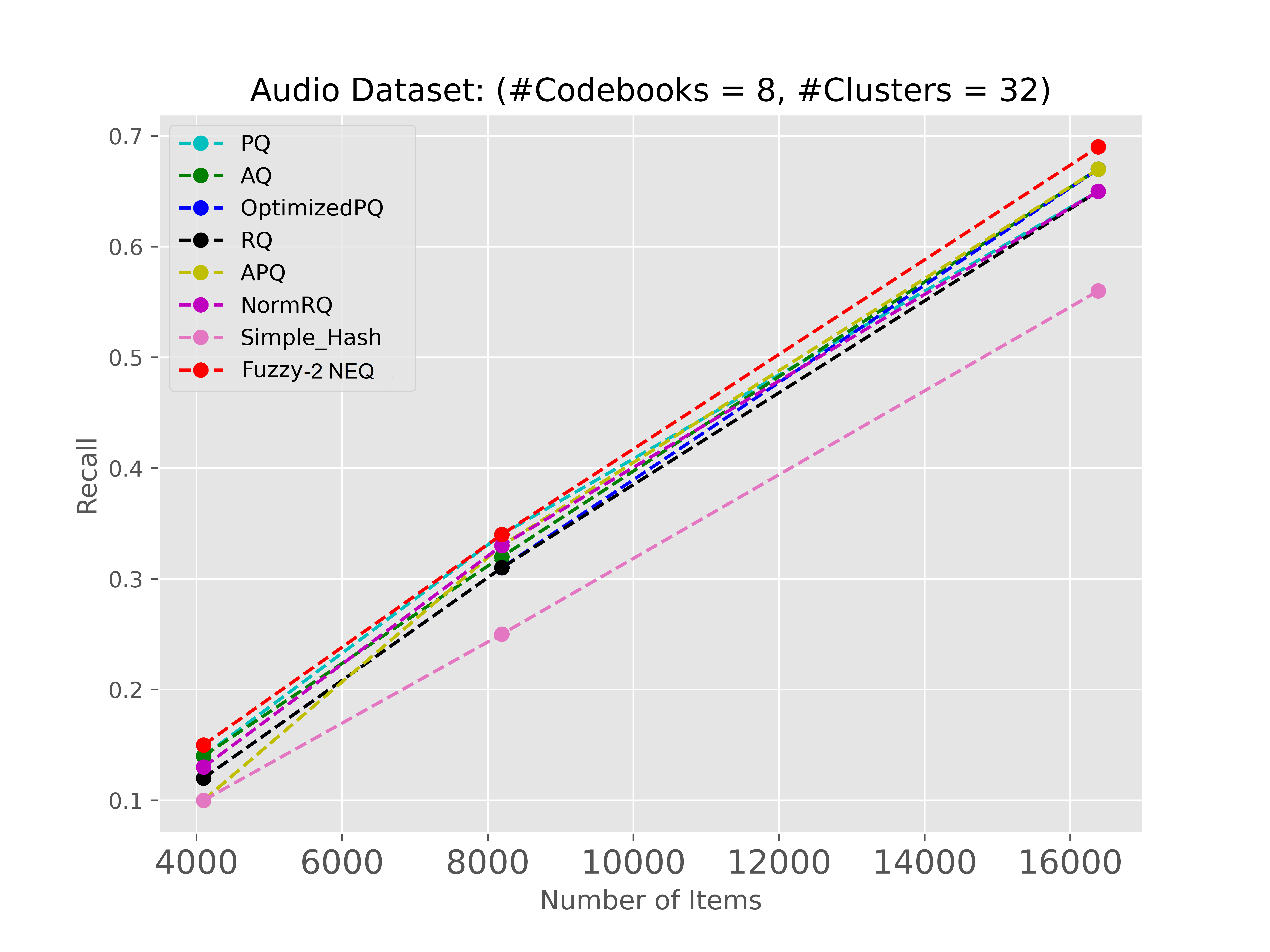}
  \caption{Audio dataset comparison plot}\label{fig:awesome_image2}
\endminipage\hfill
\minipage{0.33\textwidth}%
  \includegraphics[width=\linewidth]{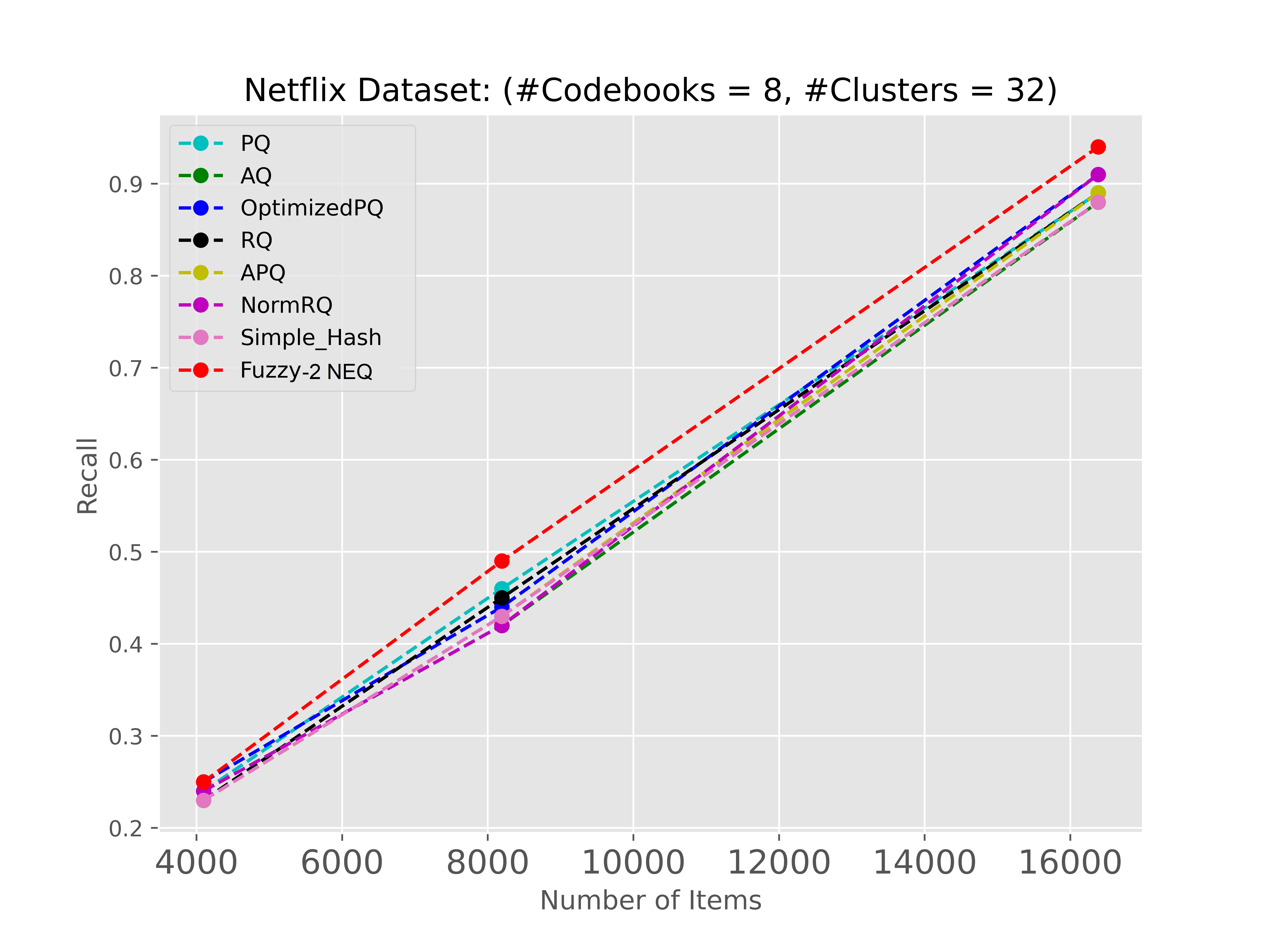}
  \caption{Netflix dataset comparison plot}\label{fig:awesome_image3}
\endminipage
\end{figure*}
Figure \ref{fig:awesome_image1}, \ref{fig:awesome_image2}, and \ref{fig:awesome_image3} presents a graphical comparison between methods PQ, AQ, OptimizedPQ, RQ, APQ, NormRQ, Simple\_Hash, and Fuzzy-2 NEQ (proposed) for 3 benchmark datasets: Cifar60k, Netflix, and Audio, which are placed left, middle, and right side respectively. In addition, figure \ref{fig:timep} presents a time performance comparison.
\subsection{Fuzzy-2 NEQ vs. NEQ}
In order to allow an easier discussion of the results, 
detailing the dimensions of datasets used in the experimentation would be useful. Table \ref{table:dataset} indicates the number of records used for the purpose of training as well as the dimension of each dataset.
The source of variation in our method would be the number of codebooks, the number of clusters, and the number of items. For each dataset, we used random sampling with replacement (e.g. bootstrapping) to select training items. To obtain a statistically unbiased result, we repeated this process 10 times for each dataset, using a fixed set of parameters. Table \ref{table:stdd} shows the average running time and average standard deviation of recall for each model on different datasets and parameter settings across the 10 iterations. In recommender systems, recall (\ref{rec}) is a key performance metric that holds great importance. This is due to its ability to the degree to which a model can recognize relevant data. Recall, defined as the proportion of relevant items that the model correctly retrieves, serves as a crucial indicator of the effectiveness of a recommender system \cite{Huilgol2020-cc}.

\begin{table}[H]
\renewcommand{\arraystretch}{0.7}
\begin{center}
\caption{Dataset Dimensions and Size of Training Set}
\begin{tabular}{ccc}
\hline
Dataset & Dimension& Training\_Size (items) \\
\toprule
Cifar60k& (512, 60,000) & 50,000 \\
Audio& (192,53387) &  40,000\\
Netflix& (300,17700) &  11,700\\
\bottomrule
\end{tabular}

\label{table:dataset}
\end{center}
\end{table}
Prior to providing the analysis of the results, it would be notable to having a summarized description of the baseline methods in this study.
\begin{itemize}
  \item PQ (Product Quantization) involves partitioning the original dataset into $M$ sub-datasets with $d/M$ features each, training independent K-means codebooks for each sub-dataset, and approximating $x$ by fusing its corresponding $d/M$-dimensional codewords \cite{zhong2022evaluating}.
  \item OPQ (Optimized Product Quantization) utilizes an orthonormal matrix $R$ to rotate dataset items before applying PQ, resulting in reduced quantization error, especially for correlated or variably distributed features, though it involves a more intricate codebook training phase with multiple rounds of optimizing codebooks and the rotation matrix $R$ \cite{guo2022semantic}.
  \item RQ (Residual Quantization) involves training codebooks where each codebook covers all features with $d$-dimensional codewords, using K-means on the original data for the first codebook and residues for subsequent codebooks in a recursive manner, improving representation through residual-based training \cite{fan2022comprehensive}.
  \item AQ (Additive Quantization) enhances RQ by collectively optimizing all $M$ codebooks, utilizing Beam search for efficient encoding and employing a least-square relation to optimize the codebooks while considering the proposed encoding for optimal codeword index determination \cite{fan2022comprehensive}.

  \item NEQ involves separately quantizing the direction and norm of a signal, demonstrating improved efficiency for maximum inner product search (MIPS) and reduced quantization error for large dynamic range data, representing an advancement in performance compared to prior methods, but it operates with discrete quantization, and removing this constraint could enhance codebook definition and quality  \cite{dai2020norm}.
  \item Norm RQ is the Norm-Explicit version of RQ which uses the concept of NEQ \cite{dai2020norm}.
  \item Simple Locality Sensitivity Hash challenges the notion that asymmetric LSH is necessary for maximum inner product search (MIPS). They present a symmetric LSH approach that rivals asymmetric LSH in performance and theoretical guarantees when considering normalized queries and bounded data vectors, suggesting that asymmetry might not be necessary or advantageous for MIPS \cite{al2021survey}.
  \item Additive Product Quantization (APQ) involves splitting a vector into $M_{1}$ orthogonal components (similar to PQ), which are then encoded by AQ into M2 bytes each. This hybrid compression approach uses $M_{1}\times M_{2}$ bytes for encoding each vector, proving to be more accurate in experiments compared to PQ compression with $M = M_{1}\times M_{2}$ .
\end{itemize}

To measure the performance of the model on a set of 4 parameters in terms of recall and precision Table \ref{table:exp} is served to fulfil this purpose.

\begin{table}[H]
\begin{center}
\caption{Average Item vs. Recall rate of items in Cifar60k, Audio and in Netflix dataset for each model with regard to the 4 types of setting in 10 iterations \protect\label{table:recall}}
\scalebox{1}{
\setlength{\tabcolsep}{1pt}
\begin{tabular}{@{}$l^l^l^l^l^l^l@{}}
\toprule
Model (\%)       & \multicolumn{2}{l}{Cifar60k} & \multicolumn{2}{l}{Audio} & \multicolumn{2}{l}{Netflix} \\ \midrule
\#Items     & 16384         & 32768        & 16384       & 32768       & 8192         & 16384        \\ \midrule
PQ   & 27.53         & 57.41         & 34.57        & 64.88        & 44.30         & 89.83         \\
OPQ   & 27.56         & 57.83         & 31.62        & 66.23        & 44.20         & 91.48         \\
AQ   & 26.40         & 56.73         & 32.22        & 67.28        & 42.40         & 88.23         \\
RQ   & 21.05         & 56.67         & 31.17        & 65.32        & 43.20         & 90.21         \\
NormRQ   & 24.10         & 56.33         & 33.41        & 66.70        & 42.94         & 91.23         \\
Simple Hash & 24.42       & 55.17         & 33.35       & 65.12        & 40.13         & 88.37\\
NEQ   & 27.52         & 56.83         & 34.35        & 66.52        & 42.97         & 91.75         \\
\rowstyle{\bfseries}Fuzzy-2 NEQ &  29.11         & 59.34         & 34.84        & 69.71        & 49.32         & 94.65         \\
APQ         & 25.52         & 56.60         & 33.19       & 64.70        & 42.95         & 89.45        \\ 
\bottomrule
\rowstyle{\bfseries}
Difference Range\% & [+2,+8]& [+2,+4] & [+0.3,+4] & [+2,+5] & [+5,+9] & [+3,+6] \\
\setlength{\tabcolsep}{-1pt}
\end{tabular}}
\end{center}

\end{table}
Adding a table with Precision and F1-score results as secondary performance measures would be notable.
\begin{table}[H]

\begin{center}
\caption{Average Precision (Prec) and F1-score (F1-s) of models on benchmark datasets \protect\label{table:pre_score}}
\scalebox{1}{
\setlength{\tabcolsep}{1pt}
\begin{tabular}{@{}$l^l^l^l^l^l^l@{}}
\toprule
Model       & \multicolumn{2}{l}{Cifar60k} & \multicolumn{2}{l}{Audio} & \multicolumn{2}{l}{Netflix} \\ \midrule
\#Items     & Prec         & F1-s        & Prec       & F1-s       & Prec         & F1-s        \\ \midrule
PQ   & 23.1         & 32.95         & 29.81        & 40.85        & 65.30         & 75.62         \\
OPQ   & 24.53         & 34.41         & 25.14        & 36.44        & 59.20         & 71.88         \\
AQ   & 21.70         & 31.39         & 26.52        & 38.04        & 62.68         & 73.30         \\
RQ   & 21.05         & 30.69         & 25.46        & 36.64        & 64.53         & 75.24         \\
NormRQ   & 24.10         & 34.05         & 28.79      & 40.22        & 61.89         & 73.74        \\
Simple Hash & 23.51       & 32.97         & 29.11      & 40.234        & 63.44         & 73.86\\
NEQ   & 26.43         & 36.08         & 28.37        & 39.78        & 65.78         & 76.62         \\
\rowstyle{\bfseries}Fuzzy-2 NEQ &  30.42         & 40.22         & 29.94        & 41.89        & 67.75         & 78.97        \\
APQ         & 21.54         & 31.20         & 27.02       & 38.12        & 64.32         & 74.83        \\ 
\bottomrule
\rowstyle{\bfseries}
Difference Range\% & [+4,+9]& [+4,+9] & [+0.01,+4] & [+1,+4] & [+2,+6] & [+2,+7] \\
\setlength{\tabcolsep}{-1pt}
\end{tabular}}
\end{center}

\end{table}

\begin{table}[H]
\begin{center}
\caption{Average (Std, time) for each model in 10 iterations \protect\label{table:stdd}}
\setlength{\tabcolsep}{3pt}
\begin{tabular}{@{}$l^l^l^l^l^l^l@{}}
\toprule
Model       & \multicolumn{2}{l}{Cifar60k (Dataset1)} & \multicolumn{2}{l}{Audio (Dataset2)} & \multicolumn{2}{l}{Netflix (Dataset3)} \\ \midrule
Average     & Time          & Std          & Time         & Std        & Time          & Std         \\ \midrule
PQ   & 25.9492         & 0.2691         & 4.1151        & 0.2740     & 7.5041         & 0.3234       \\
OPQ   & 118.1568         & 0.2654         & 26.2197        & 0.2755       & 44.2201         & 0.3270       \\
RQ   &  29.4702       & 0.2693         & 5.5542         & 0.2740      & 9.0250         & 0.3240        \\

NormRQ   & 81.0254         & 0.2677         & 18.4240        & 0.2738      & 26.1655         & 0.3242       \\

APQ   &  38.3903      & 0.2643         &     7.7638    & 0.2740      & 13.5324         & 0.3251         \\
Simple Hash   &  35.4422      & 0.2682         &     8.4322   & 0.2743       & 12.7213         & 0.3238         \\
\rowstyle{\bfseries}Fuzzy-2 NEQ & 26.3943         & 0.2671         & 4.2113        & 0.2733      & 7.1144         & 0.3225       \\
AQ   &  26.1226      & 0.2691         &     4.2375    & 0.2770       & 7.7088         & 0.3205  \\ \bottomrule
\setlength{\tabcolsep}{-3pt}
\end{tabular}

\end{center}
\end{table}
\begin{table}[H]
\begin{center}

\caption{CPU and Memory usage of Fuzzy-2 NEQ with two other baseline models on Cifar60k as the largest benchmark dataset\protect\label{table:usg}}

\begin{tabular}{@{}lll@{}}
\toprule
Models      & CPU(\%) & Memory(\%) \\ \midrule
NEQ         & 19.7\%         & 90.1\%            \\
Fuzzy-2 NEQ & 11.4\%         & 87.2\%            \\
PQ          & 20.2\%         & 89.5\%            \\ \bottomrule
\end{tabular}
\end{center}
\end{table}
\begin{table}[H]
\begin{center}
\caption{Average recall and precision during 10 iterations for each setting of Fuzzy-2 NEQ based on the number of items for different numbers of Codebooks (CB) and Clusters (Cl)} (Netflix Dataset)
\begin{tabular}{ ccccc }
\hline
\multicolumn{5}{c}{\#Items vs Average Recall \& Precision} per Iteration \\
\hline
Settings&2048 & 4096 & 8192 & 16384 \\
\hline
4-CB,4-Cl (Recall)&  0.1197 & 0.2227 & 0.4349 & 0.9188\\
8-CB,32-Cl (Recall)&  0.1379 & 0.2548 & 0.4931 & 0.9413\\
16-CB,32-Cl (Recall)& 0.1255 & 0.2542 & 0.4777 & 0.9322\\
32-CB,100-Cl (Recall)& 0.1226 & 0.2440 & 0.4831 & 0.9280\\
128-CB,100-Cl (Recall)& 0.1273 & 0.2354 & 0.4894 & 0.9433\\
\hline
4-CB,4-Cl (Precision)&  0.0077 & 0.0163 & 0.0453 & 0.6547\\
8-CB,32-Cl (Precision)&  0.0082 & 0.0178 & 0.051 & 0.6631\\
16-CB,32-Cl (Precision)& 0.0082 & 0.0181 & 0.050 & 0.6756\\
32-CB,100-Cl (Precision)& 0.0071 & 0.0177 & 0.0492 & 0.6701\\
128-CB,100-Cl (Precision)& 0.008 & 0.0182 & 0.0501 & 0.6778\\
\hline
\end{tabular}

\label{table:exp}
\end{center}
\end{table}
\begin{figure}[H]
\centerline{\includegraphics[height = 5.77cm,width=9.77cm]{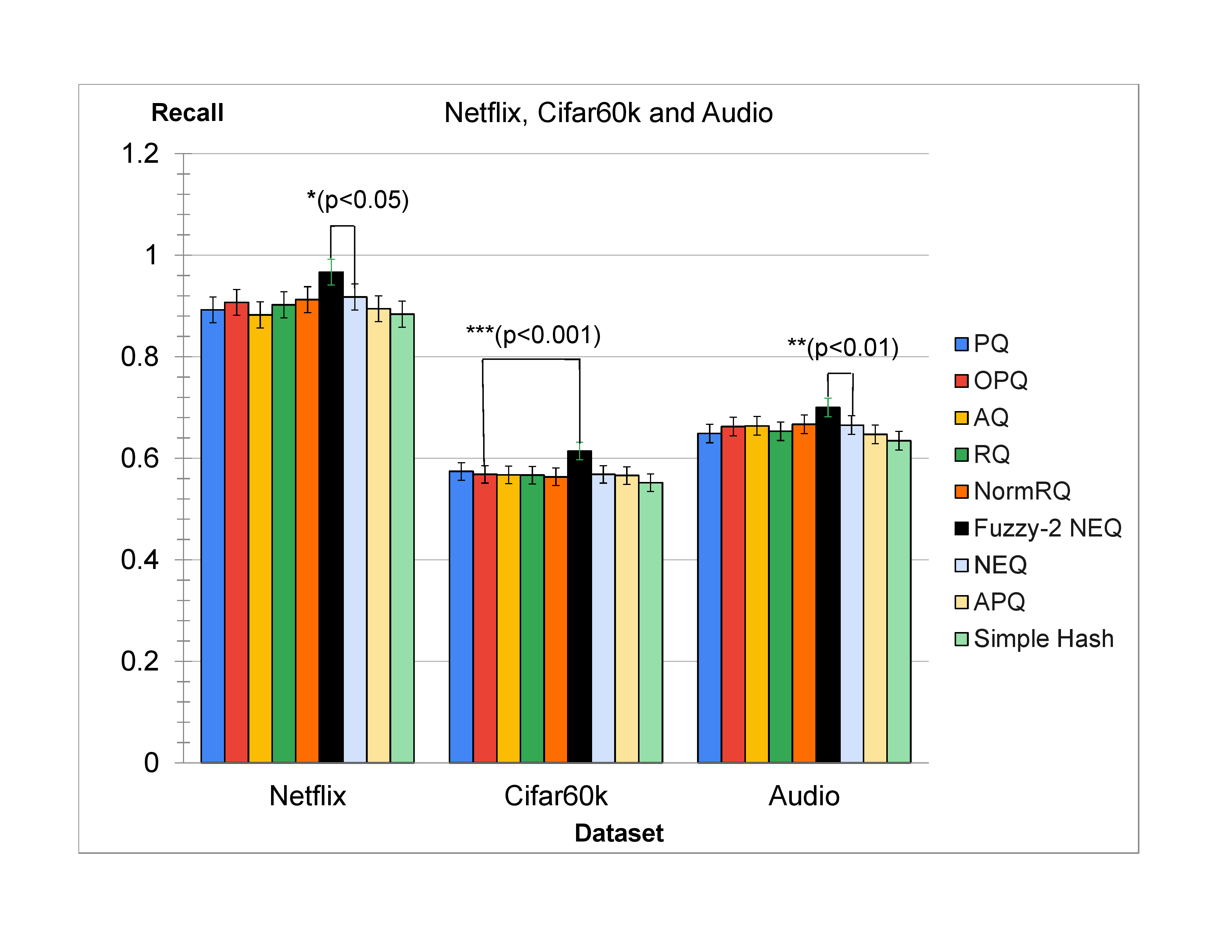}}
\caption{This figure presents a graphical comparison of the recall performance achieved by each comparison method for each dataset benchmark. Here, it can be graphically appreciated that Fuzzy-2 NEQ outperforms all methods by a salient margin.}
\captionsetup{justification=centering}
\label{fig:performancebyrecall}
\end{figure}
\subsection {Algorithmic Complexity Analysis}
A bottleneck in recommender systems is time complexity. While an accurate result is important, having a quick outcome with a high recall can still be valuable for many applications. \cite{ferrari2021troubling}. The complexity of Product Quantization is $O(\sqrt[m]{k}\times D + c)$ where $c$ is a constant referring to codebook construction cost. On the other hand, by taking a fuzzy approach and using a low-cost aggregation method, we improved the recommender system's performance in terms of recall with a negligible difference in running time. There are other techniques like matrix factorization in which the time complexity of stochastic gradient descent in these algorithms is mostly $O(k)$  \cite{gholami2021survey}. In addition, Product Quantization is a favored method for efficaciously compressing high-dimensional vectors to use around 90 per cent less memory \cite{gholami2021survey}.
\subsection {Discussion on Achievements}
There are different kinds of methods addressing the problem of finding the most interesting item or set of items for a user. However, each method has its cons and pros. Specifically, a high-performance method of tackling the problem could be using neural network-based approaches or deep learning. In this case, choosing which method could best fit the problem is related to the nature of the problem and use case. We use that as a basis to extend the notion of product quantization-based techniques in the phase of codebook construction.
Furthermore, lower time and space complexity, acceptable performance, and a technique focusing on retrieval efficiency motivate researchers and practitioners to utilize and extend this method. In summary, when the complexity of a method would be essential because of various reasons such as limitations in computational power, then a product quantization-based approach could be a suitable choice to solve the problem with the main focus on time and space complexity rather than very high expectation of accuracy in results. A table of results and recall-item curves for norm-explicit product quantization using K-means to build the codebooks have been provided to see how Norm-Explicit Quantization and the approach introduced in this research, Type 2 Fuzzy Norm-explicit Quantization (e.g. Fuzzy-2 NEQ) perform by keeping the sources of variation unchanged for both methods. Based on the comparison provided in Table \ref{table:recall}, Fuzzy-2 NEQ performs better than existing methods such as NEQ, PQ, and RQ up to +6\%, +5\%, and +8\%, respectively. A comparison bar plot, in this case, would be helpful to indicate the performance of baseline methods against Fuzzy-2 NEQ. Figure \ref{fig:performancebyrecall} presents a performance comparison of the Fuzzy-2 NEQ approach with other existing methods over approximately 15,000 records of each dataset, with respect to the number of items in the data and its recall value.

\section{Conclusion and Future Work}\label{sec:conclusion}
The importance of recommender systems in online markets, music streaming platforms, subscription streaming services (e.g., Netflix), and even social media such as YouTube or Instagram for both users and companies is tangible. To meet the market demands of every user, a recommender system with a reasonable time and space complexity and providing an acceptable percentage of relevant and accurate recommendations would be a point of interest for any business.
In conclusion, we have provided metrics such as recall and running time with a recall-item curve to see how the algorithm works in a different set of parameter settings, showing an acceptable performance during 10 iterations of each setting. By adding fuzziness to the existing Norm-Explicit Product Quantization codebook construction process, we have made progress regarding recall percentage, reflecting more relevant recommendations in output at compelling complexity. This is important since human, time, and financial resources are worth preserving and consuming carefully, and having relevant recommendations on hand could result in more success in the business in terms of finding customers interested in the product or service. Furthermore, in health recommender systems, extracting medicine based on other patients' experience has proved that applications of recommender systems in the health sector are well-worth to be researched. Eventually, using the relative norm, training the remaining codebooks is done, and we have $m$ trained codebooks. In definition, codebooks are tables mapping centroid indices (e.g., codes) to centroids and conversely \cite{noauthor_undated-au}. The resulting codebooks contain items that are compressed into pieces and then sorted to give us the top 20 preferred items. In general, the proposed method would help us to have a model which is firstly convergent, secondly considers overlapping clusters situation, and defines a more precise margin using a fusion model based on the optimality of weights. The method that was suggested has shown better results compared to quantization-based methods. In the experiments conducted on Netflix, Audio, and Cifar60k datasets, the proposed method achieved a recall of 94\%, 69\%, and 59\% respectively. This means that it was able to correctly identify a higher percentage of relevant information. Specifically, when compared to other methods such as NEQ, PQ, and RQ, the proposed method outperformed them by +6\%, +5\%, and +8\%, respectively. \newline \newline As future works, the codebook construction approach could be slightly updated, and some may change the Interval Type-2 Fuzzy Possibilistic C-Means approach, re-implemented it differently, or try to use another soft clustering method. These variations can positively impact the method's time and space complexity and, even more, may improve the model's performance. Besides, by changing the general logic of the algorithm, we can have fuzzy outputs, and by providing the user with all the prole outcomes associated with their proility, we give the user this opportunity to decide which options could excel others, which could be a source to study of user behavior patterns. Another possible action with fuzzy output could be applying some rule-mining techniques so we can choose a recommendation among the outputs \cite{kiani2022towards, kiani2019effective, andreu2011real, gupta2022gentle}. Over and above, using neural networks in the phase of codebook construction or the defuzzification process could be interesting enough to merit a study \cite{al2020deep,andreu2021single}. Fuzzification can be used in a different stage of product quantization and based on that instead of having a crisp result. As mentioned earlier, in this paper, the evidence arises that having a set of responses with a degree of truth (e.g., a fuzzy response) seems appealing to PQ-based fast recommender systems. Generally speaking, due to the massively growing number of users and items, a recommender system with competent sensitiveness, time and space complexity is a necessity to continue being intensely investigated. 

%
\appendices
\section{Optimal Codebooks and Centorids Existence Proof}
\begin{defn}[Optimal Codebook]
A codebook is called optimal either globally or locally for distribution function $F$ if it obtains the minimum value of average distortion function.
\end{defn}
\begin{thm}[Existence of k-level optimal codebooks and centroid]
\begin{proof}
There are two sufficient assumptions on distortion function $d$ to ensure the existence of the optimal codebooks and centroids as follows:\newline
A1) $d: \mathbb R^{D} \times \mathbb R^{D} \rightarrow[0, \infty)$ is continuous.
\newline
A2) For each $x, d(\tilde{x}, y) \rightarrow \infty$ as $\tilde{x} \rightarrow x$ and $\|y\|_{2} \rightarrow \infty$.\newline
Hence, for any fixed $\boldsymbol{x}, d(\boldsymbol{x}, \cdot)$ is continuous on $\left(\mathbb{\bar{R}}^{D}\right)^{k}$ by A1) and A2); hence, by average distortion equation which is 
\begin{equation} 
D(\mathcal{C}, F)=\int \min _{1 \leq j \leq k} d\left(\boldsymbol{x}, \boldsymbol{a}_{j}\right) d F(\boldsymbol{x}) \tag{A.1}
\end{equation}
and Fatou's lemma the average distortion function $D(A, F)$ is a lower semi-continuous function of the codebook $\mathcal{C}$ with respect to differential of distribution function $F(x)$. By the compactness of $\left(\mathbb{\bar{ R}^D}\right)^{k}$ the minimum value of average distortion is achieved by some Codebook $\mathcal{C}^{+}$. If $\mathcal{C}^{+}$ has some $\bar{\infty}$-valued codewords, then by A2) they can be replaced by real-valued ones. Hence an optimum $k$-level codebook exists. Observe that no restrictions need to be placed on $F$. For the above argument A1) could be generalized to lower semi-continuity in $\boldsymbol{y}$ for fixed $\boldsymbol{x}$; A2) could be generalized to
\begin{equation}
\lim _{\|\tilde{y}\| \rightarrow \infty} d(\boldsymbol{x}, \tilde{\boldsymbol{y}}) \geq d(\boldsymbol{x}, \boldsymbol{y}) \tag{A.2}
\end{equation}

for each fixed $\boldsymbol{x}$ and $\boldsymbol{y}$. If $d(\boldsymbol{x}, \boldsymbol{y})$ satisfies these conditions, then $I_{s}(\boldsymbol{x}) d(\boldsymbol{x}, \boldsymbol{y})$ does also for any Borel set $S$ in which $I_{s}$ is indicator function of the set $S$. The existence of an optimum one-level code book for this distortion function means that $\int_{s} d(\boldsymbol{x}, \boldsymbol{u}) d F(\boldsymbol{x})$ is minimized by some $\boldsymbol{u} \in \mathbb R^{D}$; that is, $S$ has a centroid. $ \square$
\end{proof}
\end{thm}
\section{}
\begin{figure}
\centerline{\includegraphics[scale=0.33]{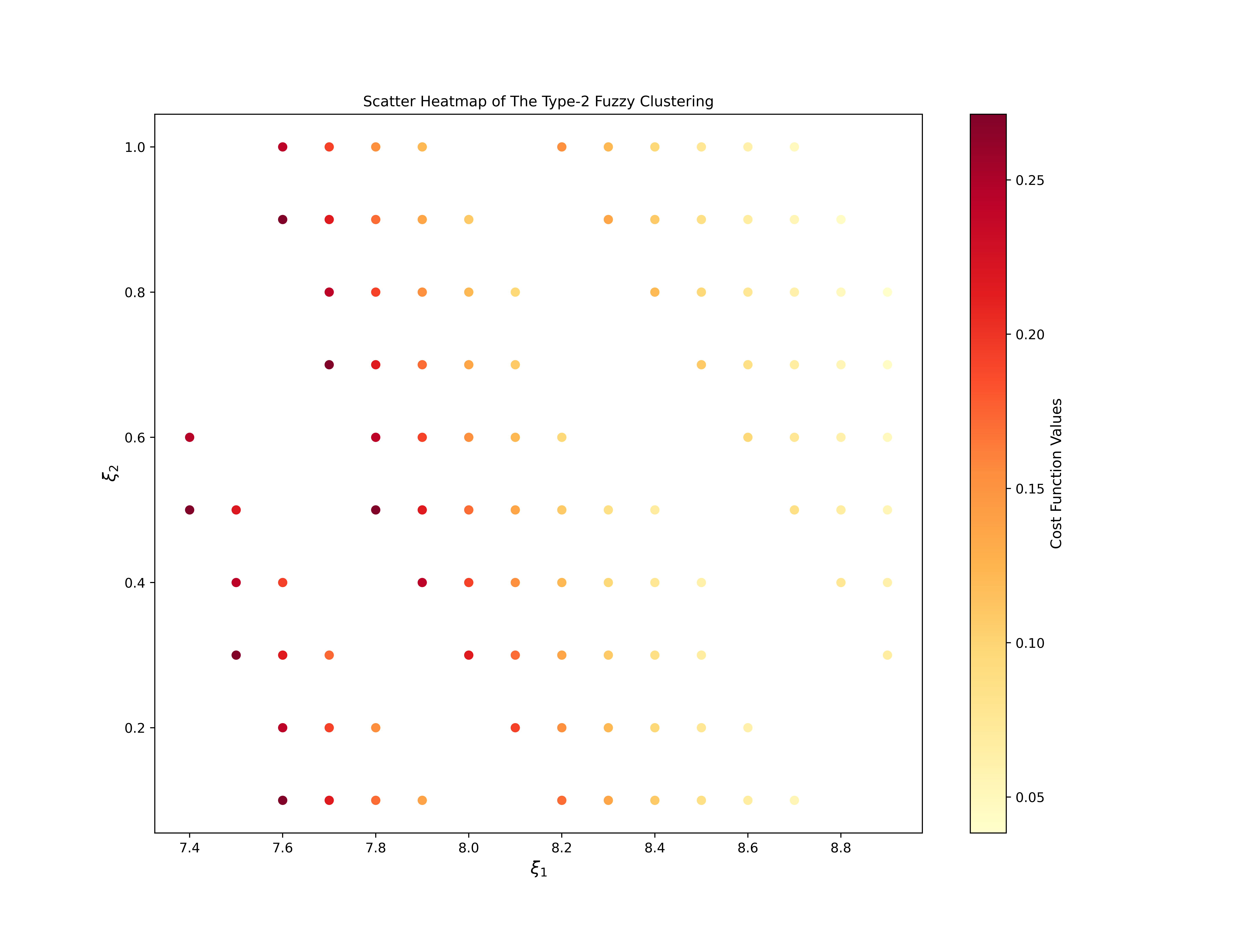}}
\caption{This figure shows a grid of cost function values for parameters $\xi_{1}$ and $\xi_{2}$ to define the T2FSs via clustering to form the codebook}
\label{fig:paremetersxi}
\end{figure}

A genetic algorithm was employed to find the optimal values for $\xi_{1}$, and $\xi_{2}$, utilizing an initial population size of 10, with a mutation rate of 0.5 and a recombination rate of 0.7. The tolerance level was set at 0.01, and the implementation was conducted using the Scipy library.
A scatter heat-map of grid values for parameters $\xi_{1}$ and $\xi_{2}$ could be intuitively used to find the optimal values for the fuzzy parameters (the colder the color, the smaller the value of the cost function). This is shown graphically in fig. \ref{fig:paremetersxi}.

\ifCLASSOPTIONcaptionsoff
  \newpage
\fi



%
\bibliography{main.bib}
\bibliographystyle{IEEEtran}

%
\begin{IEEEbiography}[{\includegraphics[width=1in,height=1.25in,clip,keepaspectratio]{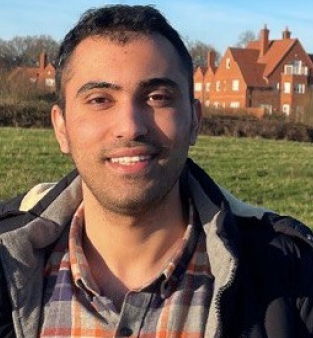}}]{Mohammadreza Jamalifard} is MSc in Data Science for the University of Essex, BSc from University of Kerman, and proleptic PhD student at the Centre for Computational Intelligence at the University of Essex. His Interests are recommendation systems, compression algorithms, quantization and machine learning.
\end{IEEEbiography}

\begin{IEEEbiography}
[{\includegraphics[width=1in,height=1.25in,clip,keepaspectratio]{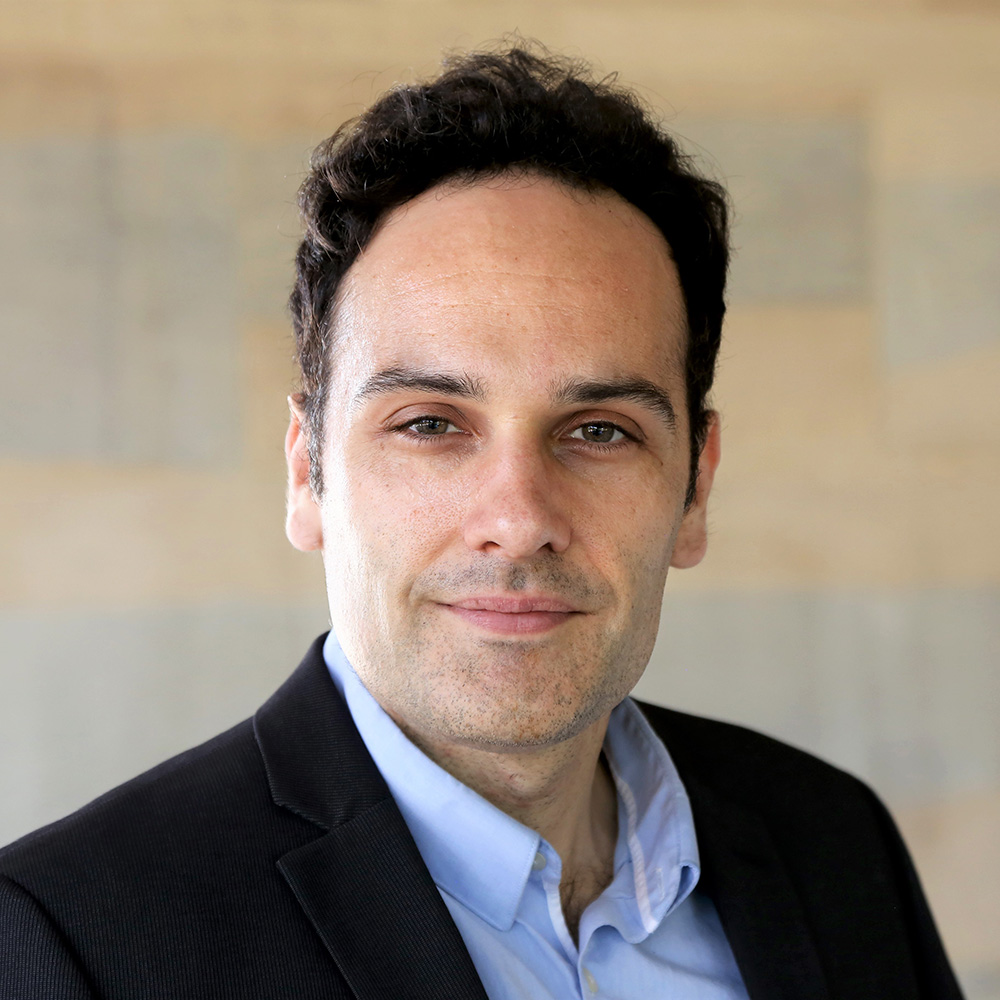}}]{Javier Andreu-Perez (SM'19)} received the Ph.D. degree, in 2012. He is a Senior Lecturer (Associate Professor, tenured) in the School of Computer Science and Electronic Engineering (CSEE) and group chair of the Smart Health Technologies Group at the University of Essex, United Kingdom. He was awarded his PhD in Intelligent Systems from Lancaster University (UK). He has expertise in artificial intelligence and fuzzy methodologies \& computing-with-words approach for uncertainty modelling in highly noisy, non-stationary, high-dimensional data. Javier's other interests are in sensor engineering, bio/neuro-engineering, and life science research. Javier has published in journals edited by Elsevier, Springer-Nature, IEEE (TFS, TSMC, TBMI, JBHI, TDCS etc.), and other venues in artificial intelligence and cognitive neuroscience. Javier's work in artificial intelligence and biomedical engineering has attracted 3600+ citations. Javier was chair of the IEEE Computational Intelligence Society (CIS) task force on Extensions of Type-1 fuzzy sets (2018-2022). Javier acts as associate Editor-in-chief of the journal Neurocomputing (Elsevier), the EUSFLAT-sponsored International Journal of Computational Intelligence Systems and other newer journals on emerging technologies. He’s been invited as an expert reviewer for significant publishers such as Science, The Lancet or BMC. He has served on the technical committee for IEEE WCCI on several occasions and is the workshop chair for the International Conference on Fuzzy Systems, 2023. He has regularly led tutorials and a special session at this event. Javier's research has been founded by UK research councils, Welcome Trust, NIHR, and IT corporations such as Nvidia, Amazon, and Oracle. Javier has an extensive portfolio of completed successful knowledge transfer and collaboration with the industry. Javier has won personal fellowships/awards for his research career from the Japan Society for the Promotion of Science (2022), Talentia Senior Fellowship from the Andalusia Scientific Council (2020) and best research associate presentation at Department of Computer Science, Imperial College London, UK (2017). 
\end{IEEEbiography}

\begin{IEEEbiography}[{\includegraphics[width=1in,height=1.25in,clip,keepaspectratio]{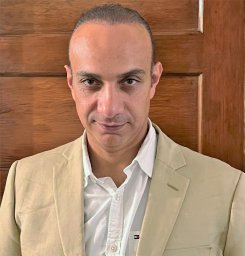}}]{Hani Hagras (Fellow'13)} is a Professor of Artificial Intelligence, Director of Impact, Director of the Computational Intelligence Centre and Head of the Artificial Intelligence Research Group, in the School of Computer Science and Electronic Engineering, University of Essex, UK. He is a Fellow of the Institute of Electrical and Electronics Engineers (IEEE'13), a Fellow of the Institution of Engineering and Technology (IET), Principal Fellow of the UK Higher Education Academy (PFHEA) and Fellow of the Asia-Pacific Artificial Intelligence Association (AAIA) His main research interests are in Explainable Artificial Intelligence (XAI) and Data Science with applications to Finance, Cyber-Physical Systems, Neuroscience, Life Sciences, Uncertainty Management, Intelligent Robotics and Intelligent Control of Industrial Processes.  He has authored more than 400 papers in international journals, conferences and books. He is among the top 2\% of the most-cited scientists in the World (Scopus August 2021). His work received funding from major research councils and industry. He holds eleven industrial patents in the field of Explainable AI. His research has won numerous prestigious international awards, where he was awarded by the IEEE Computational Intelligence Society (CIS), the 2010 Outstanding Paper Award in the IEEE Transactions on Fuzzy Systems and the 2004 Outstanding Paper Award in the IEEE Transactions on Fuzzy Systems. He was also awarded the 2015 and 2017 Global Telecommunications Business award for his joint project with British Telecom. In 2016, he was elected as Distinguished Lecturer by the IEEE Computational Intelligence Society. His work has also won best paper awards in several leading international conferences, including the 2014 and 2006 IEEE International Conference on Fuzzy Systems, the 2012 UK Workshop on Computational Intelligence and the 2016 International Conference of the BCS SGAI International Conference on Artificial Intelligence. He was awarded by the IEEE Computational Intelligence Society (CIS) the 2011 IEEE CIS Outstanding Chapter Award. In 2017,  he was awarded by the University of Essex the 2017 best Research impact award for his work with British Telecom. He acted as the Principal Investigator for a project which was awarded by the UK Technology Strategy Board, the 2011 UK Best Knowledge Transfer Partnership for London and the East Region. He also acted as the Principal Investigator for a project which was awarded the 2009 Lord Stafford Achievement in Innovation Award for East of England. In 2010, he Led a Research Students team to win First place in the RoboCup 2010. In 2007, he was Shortlisted by the Times Higher Education Supplement (THES) for the UK Young researcher of the year award. He is Associate Editor of many journals, including IEEE Transactions on Fuzzy Systems, IEEE Transactions on Artificial Intelligence, Knowledge-Based Systems, Cognitive Computations and others. He served as the General and Programme Chair of numerous major international conferences where served as the General co-Chair of the 2007 IEEE International Conference on Fuzzy Systems, and Programme Chair of the 2021 and 2017 IEEE International Conference on Fuzzy Systems as well as many other conferences.
\end{IEEEbiography}

\begin{IEEEbiography}[{\includegraphics[width=1in,height=1.25in,clip,keepaspectratio]
{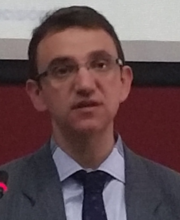}}]{Luis Martínez López (SM'18)} was born in 1970. He received the M.Sc. and Ph.D. degrees in Computer Sciences, both from the University of Granada, Spain, in 1993 and 1999, respectively. Currently, he is a Full Professor in Computer Science Department at the University of Jaén. He is the main researcher in the Research Group Intelligent Systems based on fuzzy decision analysis (Simbad2), whose main research topics are decision-making under uncertainty, decision support systems, recommender systems, sensory analysis and performance appraisal. He is also Co-Editor in Chief of the International Journal of Computational Intelligence Systems and an Associated Editor of the journals Information Fusion, and the International Journal of Fuzzy Systems and serves as a member of the journal Editorial Board of the Journal of Intelligent \& Fuzzy Systems, International Journal of Fuzzy Systems, Journal of Universal Computer Sciences and the Scientific World Journal. Additionally, he is the director of the ICT research institute at the University of Jaén. His current research interests are fuzzy decision-making, linguistic preference modelling, fuzzy systems, decision support systems, personalized marketing, computing with words and recommender systems. He co-edited eleven journal special issues on fuzzy preference modelling, soft computing, linguistic decision making and fuzzy sets theory and has been the PI in 14 R\&D projects, also published more than 120 papers in journals indexed by the SCI as well as 33 books chapters and more than 150 contributions in International Conferences related to his areas. He is a member of the European Society for Fuzzy Logic and Technology, IEEE. Co-Editor in Chief of the International Journal of Computational Intelligence Systems and an Associated Editor of the journals IEEE Transactions on Fuzzy Systems, Information Fusion, the International Journal of Fuzzy Systems, Journal of Intelligent \& Fuzzy Systems, Applied Artificial Intelligence, Journal of Fuzzy Mathematics and serves as a member of the journal Editorial Board of the Journal of Universal Computer Sciences. He received twice the IEEE Transactions on Fuzzy Systems Outstanding Paper Award in 2008 and 2012 (bestowed in 2011 and 2015 respectively). And he is Visiting Professor at the University of Technology Sydney, University of Portsmouth Isambard Kingdom Brunel Fellowship Scheme), the Wuhan University of Technology (Chutian Scholar), Guest Professor at the Southwest Jiaotong University and Honourable professor at Xihua University, both in Chengdu (China). Luis Martínez López Research has accrued 25500+ citations and an H-index of 68. Eventually, he was appointed as a Highly Cited Researcher in years 2017, 2018, 2019 and 2020 in Computer Science, According to the Web of Science. 
\end{IEEEbiography}



\end{document}